\ifx\XeTeXversion\undefined
\pdfoutput=1
\fi
\documentclass[10pt,onecolumn,aps,prx,superscriptaddress,nofootinbib,nobibnotes,floatfix]{revtex4-2}
\ifx\XeTeXversion\undefined\usepackage[T1]{fontenc}\fi
\usepackage{amsmath,amssymb,bm}
\usepackage{graphicx}
\usepackage{booktabs}
\usepackage{xcolor}
\usepackage{hyperref}
\usepackage{microtype}
\usepackage{url}
\makeatletter
\let\auto@bib@innerbib\@empty
\makeatother

\hypersetup{colorlinks=true,linkcolor=blue!60!black,citecolor=blue!60!black,urlcolor=blue!60!black}

\newcommand{\Cp}{C_p}

\newcommand{\Dfour}{\Delta^2_{\mathrm{SC,4b}}}

\newcommand{\Mfull}{M^{2,\mathrm{full}}_{\mathrm{str,4p}}}

\newcommand{\mt}{\,\mathrm{m}t}

\newcommand{\eps}{\epsilon_{\mathrm{spin}}}
\newcommand{\chivtx}[1]{\chi^{\mathrm{vtx}}_{\mathrm{SC}}(R{\le}#1)}
\newcommand{\Ekin}{E_{\mathrm{kin}}}

\begin{document}

\title{Multi-reference variational states resolve stripe selection and local \texorpdfstring{$d$}{d}-wave pairing in the Hubbard model}

\author{Honghui Shang}
\email{shanghui.ustc@gmail.com}
\author{Jinlong Yang}
\email{jlyang@ustc.edu.cn}
\affiliation{State Key Laboratory of Precision and Intelligent Chemistry, University of Science and Technology of China, Hefei 230026, China}
\date{\today}

\begin{abstract}
Whether stripes compete with $d$-wave superconductivity or promote pairing through fluctuations remains unsettled in the two-dimensional Hubbard model. Using multi-reference neural variational auxiliary-field quantum Monte Carlo (MR-nVAFQMC) at $U/t=8$, we superpose unpaired Slater determinants across stripe periods and domain-wall positions, reaching variational energies competitive with leading benchmarks on periodic lattices up to $32\times32$ sites. For cuprate next-nearest-neighbor hopping ($t'/t=-0.2$), diagonal hopping preserves electron-doped antiferromagnetism while kinetically frustrating hole motion into Yamada stripes with a dome of short-range connected $d$-wave pairing. At $1/8$ hole doping, hole delocalization across antiphase spin boundaries outweighs on-site repulsion, splitting the filled stripes of the pure model into half-filled walls with strong local pairing. Coherently delocalizing shifted walls strongly suppresses static charge modulation yet leaves short-range pairing unchanged, showing that local pairing is intrinsic to half-filled walls rather than driven by stripe fluctuations. Across intervening antiferromagnetic domains, however, pairing correlations become unresolved at one stripe period and beyond. Both parameter transfer to larger lattices without reoptimization and deep benchmark checkpoints retain stripe spin order without resolving long-range pairing, within uncertainties allowing weak superconductivity. Half-filled antiphase walls thus generate local $d$-wave pairs, whereas suppressing static stripe order alone yields no resolved coherence across stripes in the states studied.
\end{abstract}

\maketitle

\section{Introduction}
\label{sec:intro}

How unidirectional spin and charge stripes interact with $d$-wave superconductivity has been a central question in the physics of high-$T_c$ cuprates for more than three decades \cite{Tranquada1995,Fradkin2015,Keimer2015}. In $\mathrm{La}_{2-x}\mathrm{Sr}_x\mathrm{CuO}_4$, neutron scattering reveals incommensurate magnetic peaks displaced from $(\pi,\pi)$ by $\epsilon_{\mathrm{spin}}$, which grows approximately linearly with hole concentration ($\epsilon_{\mathrm{spin}}\approx\delta$, the Yamada relation) up to $\delta\approx 1/8$ before saturating at higher doping \cite{Yamada1998}. The suppression of bulk superconductivity near $1/8$ doping in $\mathrm{La}_{2-x}\mathrm{Ba}_x\mathrm{CuO}_4$ is the longstanding \emph{$1/8$ anomaly} \cite{Moodenbaugh1988}. In $\mathrm{La}_{1.875}\mathrm{Ba}_{0.125}\mathrm{CuO}_4$, static period-four charge order ($\lambda_c=4$) and period-eight antiphase spin order ($\lambda_s=8$) coexist with a strongly depressed bulk three-dimensional (3D) transition at $T_c\approx 4\text{ K}$, whereas transport and magnetization measurements indicate two-dimensional (2D) superconducting fluctuations below about $40\text{ K}$ and a Berezinskii--Kosterlitz--Thouless transition near $16\text{ K}$ \cite{Li2007}. A proposed explanation for the suppressed interlayer coherence is an intra-plane pair-density wave (PDW), whose pairing phase alternates across adjacent stripes and whose stripe direction rotates by $90^\circ$ between layers, frustrating interlayer Josephson coupling \cite{Berg2007,Fradkin2015}. Because unidirectional antiphase stripes of charge period $\lambda_c$ and linear hole filling $f=\delta\lambda_c$ carry magnetic incommensurability $\epsilon_{\mathrm{spin}}=\delta/(2f)$, the observed slope $\epsilon_{\mathrm{spin}}=\delta$ points directly to half-filled domain walls ($f=1/2$) rather than filled walls ($f=1$, $\epsilon_{\mathrm{spin}}=\delta/2$). Whether a single-layer 2D Hubbard model at $\delta=1/8$ develops phase coherence across half-filled stripes---as uniform $d$-wave or antiphase PDW order---or supports only short-range intra-wall pairing remains a central microscopic question \cite{Emery1997,Kivelson1998,Fradkin2015,Chen2026}.

Whether the single-band 2D repulsive Hubbard model captures these intertwined phenomena has been investigated intensively with modern many-body methods \cite{Anderson1987,ZhangRice1988,Scalapino2012,Arovas2022,Qin2022,LeBlanc2015}. For the nearest-neighbor model ($t'=0$) at $\delta=1/8$ and $U/t=8$, combined density-matrix renormalization group (DMRG), auxiliary-field quantum Monte Carlo (AFQMC), density-matrix embedding, and tensor-network calculations established an insulating filled-stripe ground state ($\lambda_c=8,\lambda_s=16$) without long-range superconductivity \cite{Zheng2017,Qin2020,Sorella2023}. Negative next-nearest-neighbor hopping $t'/t<0$, relevant to cuprate Fermi surfaces and electron--hole asymmetry \cite{Andersen1995,Pavarini2001,Hirayama2018}, favors shorter stripe wavelengths; at $\delta=1/8$, infinite projected entangled-pair states find period-four stripes with suppressed superconductivity \cite{Ponsioen2019}. Pairing also depends sensitively on cylinder width: four-leg calculations find algebraically decaying pairing at finite $t'$ \cite{Jiang2019}, whereas eight-leg calculations find extremely short-ranged superconducting correlations for $t'\le 0$ \cite{JiangKivelson2026}. In a distinct setting, periodically modulating nearest-neighbor hopping enhances superconducting correlations on four- and six-leg cylinders \cite{JiangKivelson2022}. Finite-temperature simulations reveal fluctuating stripes and intertwined spin, charge, and pair correlations \cite{Huang2017,Mai2022}. Complementary cold-atom experiments observe precursors of individual stripes in mixed-dimensional settings \cite{Bourgund2025} and a pseudogap in a two-dimensional Fermi--Hubbard simulator \cite{Kendrick2026}.

Determining the 2D zero-field behavior at $\delta=1/8$ with $t'\ne 0$ has nonetheless proved delicate because distinct uniform, superconducting-stripe, and non-superconducting-stripe states compete on the milli-$t$ ($10^{-3}t$) energy scale per site and respond sensitively to boundary conditions, applied fields, and variational bias. Combining DMRG with constrained-path AFQMC under twist-averaged boundary conditions and applied pairing fields---taking the thermodynamic limit at fixed pairing field before extrapolating the field to zero---Xu \emph{et al.} reported $d$-wave superconductivity coexisting with partially filled stripes ($f\approx 0.6$--$0.8$) on the hole-doped side and N\'eel antiferromagnetism on the electron-doped side \cite{Xu2024}. Working directly at zero external field on periodic 2D tori with neural quantum states \cite{Carleo2017,Luo2019,Wu2024}, Rende \emph{et al.} reported finite extrapolated $d$-wave pairing up to $24\times24$ using symmetry-preserving BCS backflow \cite{Rende2026}. Extending Pfaffian backflow \cite{Roth2025,Chen2026pnas} and neural-network fermionic projected entangled pair states (NN-fPEPS) up to $32\times32$ tori, Chen \emph{et al.} reached lower variational energies and revealed a near-degeneracy between a nearly uniform $d$-wave state and lower-energy striped states carrying very weak or statistically unresolved long-range superconductivity \cite{Chen2026}. On $8\times8$, Viteritti \emph{et al.} found that near-degenerate determinant and Pfaffian backflow states can exhibit different pair correlations, but that symmetry restoration and variance reduction bring the three wavefunction families toward a common picture of coexisting stripe and pairing correlations on this finite lattice \cite{Viteritti2026}. Related neural backflow calculations show how symmetry projection and multi-determinant improvements change both energies and correlation functions \cite{Loehr2025}.

Two methodological considerations motivate probing this competing regime with multi-reference unpaired auxiliary-field wavefunctions. First, a single-reference variational state---even after space-group projection---inherits a single reference stripe period and rigid inter-wall spacing, preventing inequivalent stripe periods and shifted domain-wall arrangements from competing or fluctuating coherently within one quantum state. Second, variational families differ in how pairing enters their reference states. Canonical number-projected BCS and Pfaffian wavefunctions conserve total particle number just as Slater determinants do, but embed explicit pairing fields into their reference geminals; by contrast, each individual unprojected particle-conserving Slater determinant $|\Phi_k\rangle$ carries no explicit pairing field and has vanishing connected two-body pairing ($\Cp=0$), whereas space-group projection, multi-determinant superposition, and the shared imaginary-time auxiliary-field projector generate non-Wick four-fermion correlations without geminal pairing initialization. Although a variational upper bound on total energy implies neither upper nor lower bounds on order parameters, reaching competitive energies from particle-conserving Slater pools provides a complementary test of which features of stripe order govern local pair formation and which limit long-distance coherence.

\begin{figure}[t!]
\centering
\includegraphics[width=0.96\textwidth]{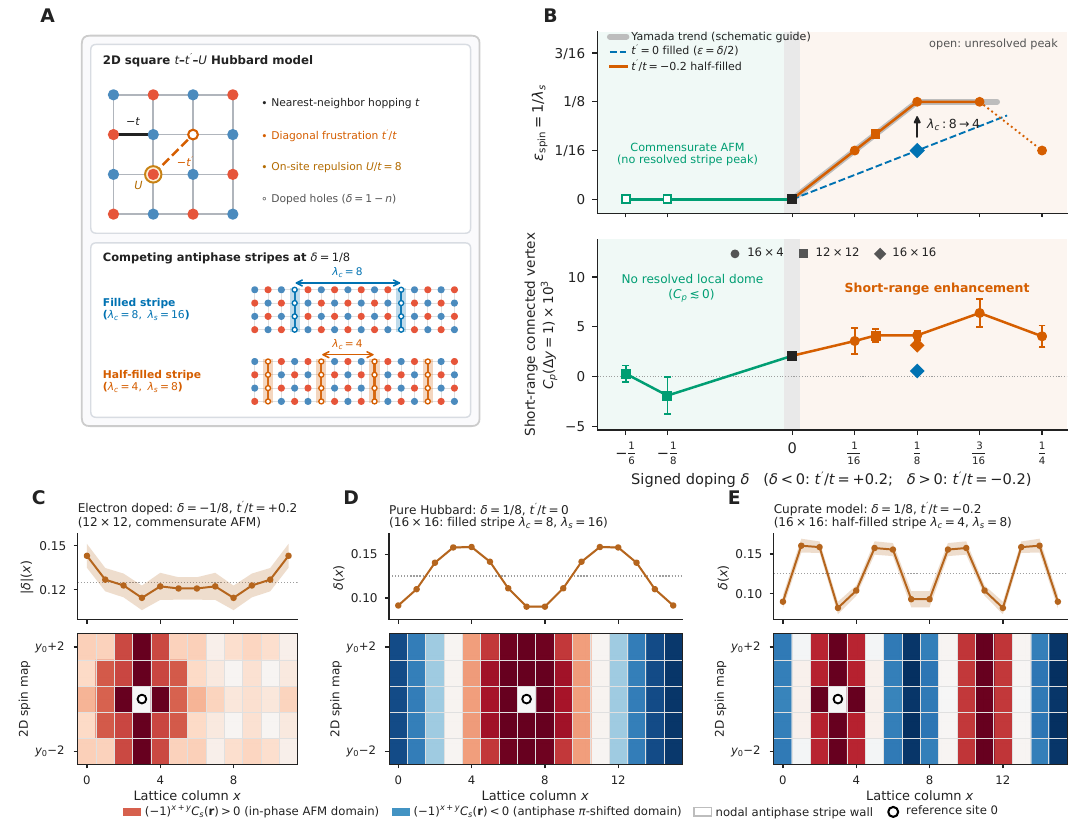}
\caption{\textbf{Square-lattice $t$--$t'$--$U$ Hubbard model, Yamada relation, and real-space stripe correlations across doping.}
(\textbf{A}) Schematic of the 2D square $t$--$t'$--$U$ Hubbard model ($U/t=8$; top) and competing filled ($\lambda_c=8,\lambda_s=16$) versus half-filled ($\lambda_c=4,\lambda_s=8$) antiphase stripe orders at $\delta=1/8$ (bottom).
(\textbf{B}) Spin incommensurability $\eps=1/\lambda_s$ (top) and nearest-neighbor connected $d$-wave pairing vertex $\Cp(\Delta y=1)$ (bottom) versus signed doping $\delta$ ($\delta<0$: $t'/t=+0.2$; $\delta>0$: $t'/t=-0.2$) on $16\times4$ (circles), $12\times12$ (squares), and $16\times16$ at $\delta=1/8$ (diamonds: blue, $t'=0$; vermilion, $t'/t=-0.2$, coinciding with $\eps=1/8$ in the top panel). Top: open symbols denote spin peaks below $3\sigma$; the grey band guides the Yamada trend ($\eps=\delta$ up to $\delta=1/8$ followed by an $\eps=1/8$ plateau across $\delta\le 3/16$), and the dashed blue line marks filled stripes ($\eps=\delta/2$).
(\textbf{C}--\textbf{E}) Intrinsic 1D carrier-density profiles $\delta(x)$ [or $|\delta|(x)$ in (\textbf{C}); top] and five-row 2D staggered spin-correlation maps $(-1)^{x+y}C_s(\mathbf r)$ around a reference site (open circle; bottom) for (\textbf{C}) electron-doped $12\times12$ at $\delta=-1/8$ ($t'/t=+0.2$, commensurate AFM), (\textbf{D}) $16\times16$ at $\delta=1/8$, $t'=0$ (filled stripe, $\lambda_c=8,\lambda_s=16$), and (\textbf{E}) $16\times16$ at $\delta=1/8$, $t'/t=-0.2$ (half-filled stripe, $\lambda_c=4,\lambda_s=8$). Error bars denote 1 s.e.}
\label{fig:phase}
\end{figure}

Here we develop multi-reference neural variational auxiliary-field quantum Monte Carlo (MR-nVAFQMC) to study the 2D square-lattice $t$--$t'$--$U$ Hubbard Hamiltonian (Fig.~\ref{fig:phase}A) on periodic $L_x\times L_y$ tori ($N=L_xL_y$ sites),
\begin{equation}
\hat H=-t\sum_{\langle ij\rangle,\sigma}(\hat c^\dagger_{i\sigma}\hat c_{j\sigma}+\mathrm{H.c.})
-t'\sum_{\langle\langle ij\rangle\rangle,\sigma}(\hat c^\dagger_{i\sigma}\hat c_{j\sigma}+\mathrm{H.c.})
+U\sum_i\hat n_{i\uparrow}\hat n_{i\downarrow},
\label{eq:ham}
\end{equation}
where $\langle ij\rangle$ and $\langle\langle ij\rangle\rangle$ denote nearest- and next-nearest-neighbor pairs, $\hat n_{i\sigma}=\hat c^\dagger_{i\sigma}\hat c_{i\sigma}$, and we set $t=1$ with $U/t=8$. Hole doping $\delta=1-N_e/N$ is studied in the spin-balanced canonical sector $N_\uparrow=N_\downarrow=N_e/2$ for both $t'=0$ and cuprate hopping $t'/t=-0.2$; physical electron doping ($n=1+|\delta|$ at $t'/t=-0.2$) maps via the bipartite particle--hole transformation $\hat c_{i\sigma}\to(-1)^{x_i+y_i}\hat c^\dagger_{i\sigma}$ to filling $1-|\delta|$ at reversed diagonal hopping $t'/t=+0.2$, displayed at signed doping $\delta<0$.

\section{Results}
\label{sec:results}

\subsection{Doping asymmetry, the Yamada relation, and kinetic stabilization of half-filled stripes}
\label{sec:yamada}

Figure~\ref{fig:phase}B--E traces how magnetic order and local $d$-wave pairing evolve across electron and hole doping in the absence of external pinning fields. On the electron-doped branch ($\delta<0$, $t'/t=+0.2$), doped carriers move coherently on the same magnetic sublattice without nucleating antiphase domain walls: the staggered spin correlation $(-1)^{x+y}C_s(\mathbf r)$ remains uniformly positive across the $12\times12$ torus at $\delta=-1/8$ (Fig.~\ref{fig:phase}C), keeping the magnetic Bragg peak locked at the commensurate N\'eel wavevector $(\pi,\pi)$ ($\eps=0$; Fig.~\ref{fig:phase}B, top), while no local connected $d$-wave pairing dome develops (Fig.~\ref{fig:phase}B, bottom). On the hole-doped branch ($\delta>0$, $t'/t=-0.2$), by contrast, the magnetic peak splits into incommensurate pairs $(\pi\pm 2\pi\eps,\pi)$ whose shift follows the half-filled Yamada relation $\eps=\delta$ across commensurate lattices at $\delta=1/16$ ($\lambda_s=16$), $\delta=1/12$ ($\lambda_s=12$), and $\delta=1/8$ ($\lambda_s=8$). Above $\delta=1/8$ on $16\times4$, the magnetic wavevector saturates at the $\eps=1/8$ plateau at $\delta=3/16$---raising the linear wall filling from $f=1/2$ to $f=3/4$ while the stripe spin peak weakens nearly threefold---before departing from the $\eps=1/8$ plateau to a weaker $\eps=1/16$ peak at heavy overdoping ($\delta=1/4$), where static charge modulation falls below $3\sigma$. Across the entire hole-doped range $\delta\in[1/16,1/4]$, the nearest-neighbor connected $d$-wave pairing vertex $\Cp(\Delta y=1)$ (Section~\ref{sec:method}) forms a broad positive dome, reaching its maximum around $\delta=3/16$ where period-eight spin stripe order begins to soften (Fig.~\ref{fig:phase}B, bottom).

\begin{figure}[ht!]
\centering
\includegraphics[width=0.96\textwidth]{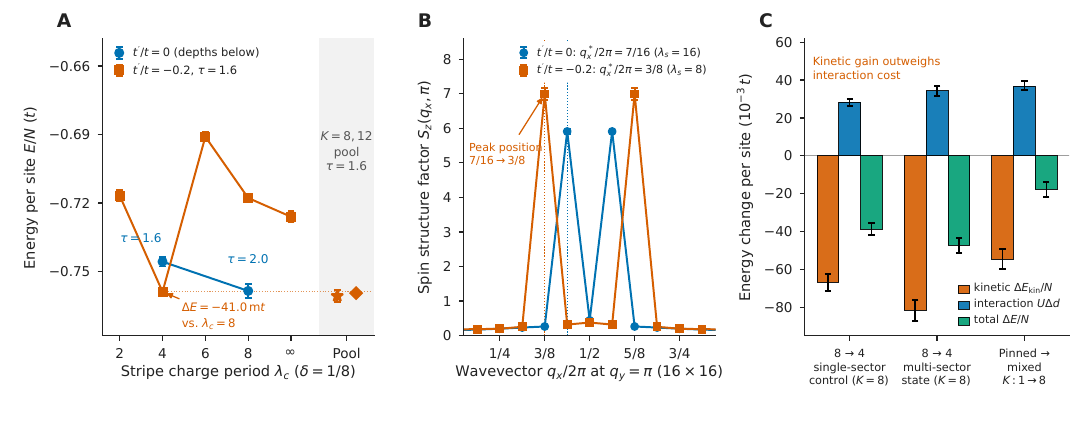}
\caption{\textbf{Stripe wavelength switch and kinetic stabilization at $\delta=1/8$.}
(\textbf{A}) Ground-state energy per site $E/N$ on $16\times4$ across charge periods $\lambda_c\in\{2,4,6,8,\infty\}$ for $t'/t=0$ (blue circles) and $t'/t=-0.2$ (vermilion squares, $\tau=1.6$), alongside unbiased $K=8$ RR-CV and $K=12$ multi-period pools (shaded column).
(\textbf{B}) Longitudinal spin structure factor $S_z(q_x,\pi)$ on $16\times16$ at $\delta=1/8$, shifting from $q_x^*/2\pi=7/16$ ($\lambda_s=16$, $t'=0$) to $3/8$ ($\lambda_s=8$, $t'/t=-0.2$).
(\textbf{C}) Kinetic ($\Delta\Ekin/N$), interaction ($U\Delta d$), and total ($\Delta E/N$) energy changes on $16\times4$ ($t'/t=-0.2$, $\tau=1.6$) from $\lambda_c=8$ to $\lambda_c=4$ single-sector control (left), $\lambda_c=8$ to $\lambda_c=4$ multi-sector $K=8$ (middle), and pinned $K=1$ to multi-sector $K=8$ at fixed $\lambda_c=4$ (right). Error bars denote 1 s.e.}
\label{fig:wave}
\end{figure}

To identify how negative $t'/t$ selects half-filled Yamada stripes over the filled stripes of the pure Hubbard model, we compare candidate charge periods $\lambda_c\in\{2,4,6,8,\infty\}$ at $\delta=1/8$ in Fig.~\ref{fig:wave}. On the four-leg $16\times4$ torus, whose length $L_x=16$ accommodates both $\lambda_s=16$ and $\lambda_s=8$ at deep projection ($\tau=1.6$), the filled $\lambda_c=8$ stripe ($f=1$) is favored at $t'=0$ whereas introducing $t'/t=-0.2$ reverses the energetic ordering (Fig.~\ref{fig:wave}A): the half-filled $\lambda_c=4$ manifold ($f=1/2$) becomes the clear variational minimum among all single-period sectors, lying $(41.0\pm1.1)\mt$ per site below $\lambda_c=8$ and well below $\lambda_c=2$, $\lambda_c=6$, and the uniform ($\lambda_c=\infty$) state. When all candidate periods compete simultaneously inside unbiased $K=8$ and $K=12$ multi-period pools (Fig.~\ref{fig:wave}A, shaded column), the optimized wavefunctions spontaneously select the $\lambda_s=8$ ($\lambda_c=4$) sector and match or slightly improve upon the independently optimized $\lambda_c=4$ energy.

The same period-halving transition occurs on the isotropic 256-site $16\times16$ torus at $\delta=1/8$: the Bragg peak of $S_z(q_x,\pi)$ shifts from $q_x^*/2\pi=7/16$ ($\lambda_s=16,\lambda_c=8$) at $t'=0$ to $3/8$ ($\lambda_s=8,\lambda_c=4$) at $t'/t=-0.2$ (Fig.~\ref{fig:wave}B), replacing two broad filled walls with four half-filled antiphase walls in real space (Fig.~\ref{fig:phase}D,E) and increasing the intra-wall nearest-neighbor pairing vertex more than fivefold. Decomposing the total energy into hopping and on-site repulsion, $E/N=\Ekin/N+Ud$ with double occupancy $d=\langle\hat n_\uparrow\hat n_\downarrow\rangle$ (Fig.~\ref{fig:wave}C), reveals why half-filled stripes win: splitting filled $\lambda_c=8$ walls into twice as many half-filled $\lambda_c=4$ walls increases local double occupancy and raises the Hubbard interaction energy ($U\Delta d>0$), yet this repulsion penalty is outweighed more than twofold by the gain in transverse kinetic energy ($\Delta\Ekin/N<0$) as holes hop across closer antiphase domain boundaries without leaving extended strings of overturned spins. Sublattice kinetic frustration thus unifies both sides of the phase diagram: positive effective $t'/t$ on the electron-doped side assists intrasublattice motion within a uniform N\'eel background, whereas negative $t'/t$ frustrates intrasublattice hopping on the hole-doped side and stabilizes half-filled antiphase walls.

\subsection{Controlled stripe superpositions: pinned and delocalized domain walls pair alike}
\label{sec:fluctuating}

Testing whether transverse stripe fluctuations amplify $d$-wave pairing requires separating three situations that are easily conflated: (i) a \emph{pinned stripe}, whose domain walls sit at fixed lattice columns with static hole-density modulation $A_4>0$; (ii) an \emph{incoherent translation mixture} of shifted copies of that pinned stripe, which sets the one-body modulation to $A_4=0$ and transfers its Bragg weight $\frac{N}{4}A_4^2$ into the connected density structure factor $N_c(\pi/2,0)$ while leaving every linear translation-averaged observable---total energy $E/N$, full $S_{cc}(\mathbf q)=\frac{N}{4}A_4^2\delta_{\mathbf q,(\pm\pi/2,0)}+N_c(\mathbf q)$, bare $P_d(\mathbf r)$, and bare $d$-wave Bloch eigenvalues---identically unchanged [whereas Wick-subtracted $\Cp(\mathbf r)=P_d(\mathbf r)-N_0[\bar G](\mathbf r)$ shifts slightly only because $\bar G$ is translation-averaged before squaring]; and (iii) a \emph{coherent quantum superposition} of shifted stripe sectors, which differs from the classical mixture solely through off-diagonal quantum interference between distinct domain-wall positions.

\begin{figure}[ht!]
\centering
\includegraphics[width=0.96\textwidth]{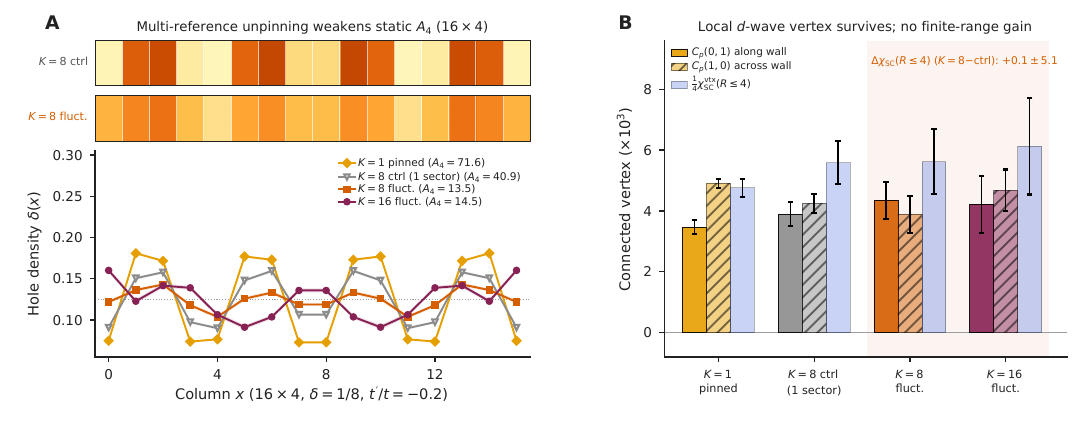}
\caption{\textbf{Multi-reference stripe unpinning and local $d$-wave pairing at $\delta=1/8$, $t'/t=-0.2$.}
(\textbf{A}) Intrinsic 2D hole-density maps comparing the $K=8$ single-sector control and fluctuating multi-sector $K=8$ state on $16\times4$ ($\tau=1.60$; top), together with 1D hole profiles $\delta(x)$ and period-four amplitudes $A_4\times10^3$ across $K=1$ pinned, $K=8$ control, $K=8$ fluctuating, and $K=16$ fluctuating states (bottom).
(\textbf{B}) Nearest-neighbor connected $d$-wave vertex along the walls [$\Cp(0,1)$], across the walls [$\Cp(1,0)$], and scaled cumulative short-range sum $\frac14\chivtx{4}$ across the $K=1\to 16$ unpinning series, showing surviving local $d$-wave pairing without finite-range enhancement. Error bars denote 1 s.e.}
\label{fig:mech}
\end{figure}

Figure~\ref{fig:mech} carries out this controlled comparison on $16\times4$ at fixed Hamiltonian and projection depth ($\delta=1/8$, $t'/t=-0.2$, $\tau=1.60$), contrasting a pinned single-reference stripe ($K=1$), a single-sector $K=8$ control localized in one wall sector, and delocalized multi-sector $K=8$ and $K=16$ superpositions distributed across translated wall positions. Relative to the matched single-sector $K=8$ control, coherent multi-sector mixing suppresses the intrinsic-frame period-four hole amplitude $A_4$ roughly threefold while shifting weight into the connected charge peak $N_c(\pi/2,0)$ at conserved total $S_{cc}(\pi/2,0)$ (Fig.~\ref{fig:mech}A), and re-optimizing the projector under coherent shifted-stripe superposition lowers the variational energy by $\sim 3\mt$ per site below the pinned state.

Across the entire $K=1\to 16$ unpinning series, short-range connected $d$-wave pairing remains robust and essentially unchanged (Fig.~\ref{fig:mech}B). Subtracting the normal Wick contraction $N_0(\mathbf r)$ from $P_d(\mathbf r)$ yields a positive nearest-neighbor connected vertex along the walls in all four states; by contrast, matched-depth uniform paramagnetic and filled $\lambda_c=8$ reference controls at identical Hamiltonian and projector architecture collapse both variational energy and local $\Cp(0,1)$ toward zero, confirming that the half-filled antiphase wall geometry itself generates the local pairing vertex. While expanding from $K=1$ to $K\ge 8$ restores local bond isotropy between $\Cp(1,0)$ and $\Cp(0,1)$, this restoration is already present in the pinned $K=8$ single-sector control; comparing the delocalized $K=8$ superposition directly against its matched single-sector control reveals no statistically resolved change in either the cumulative short-range vertex $\chivtx{4}$ or its strictly non-overlapping bond subset where three-site contact terms vanish identically. At inter-stripe separations ($\Delta x=\lambda_c=4$) and across distant bond pairs, connected pair correlations remain unresolved within uncertainty, and split-block cross-validated Rayleigh quotients of the full $2N\times2N$ bond-singlet pair matrix $\Gamma_{bb'}=\langle\hat b_b^\dagger\hat b_{b'}\rangle$ on $64$-site ($16\times4$ and $8\times8$) tori remain non-extensive ($\lambda_1^{\mathrm{CV}}/N\approx 8\text{--}9\times10^{-3}$) with comparable connected weight in uniform $\mathbf q=(0,0)$ and period-eight $(\pi/4,0)$ modes. Local $d$-wave pairing is therefore already an intrinsic property of an individual half-filled domain wall, whereas unpinning the charge profile within our multi-reference subspace neither amplifies local pairing nor induces resolved inter-stripe coherence.

\subsection{Ground-state energy benchmarks on large two-dimensional square tori}
\label{sec:benchmarks}

\begin{figure}[ht!]
\centering
\includegraphics[width=0.98\textwidth]{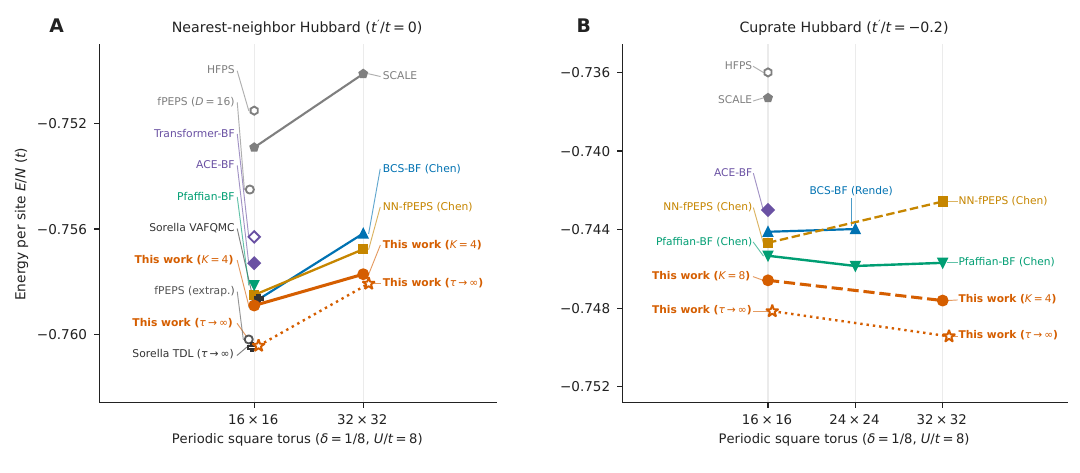}
\caption{\textbf{Ground-state energy benchmarks across large periodic 2D square tori at $\delta=1/8$ ($U/t=8$).}
Filled or solid-outlined symbols denote strictly variational energies; open symbols denote empirical $\tau\to\infty$ or $D\to\infty$ extrapolations. Because Monte Carlo standard errors span two orders of magnitude across sign-free VMC and finite-sign VAFQMC, error bars are omitted from the plot and listed in Table~\ref{tab:bench}.
(\textbf{A}) Energy per site $E/N$ on $16\times16$ and $32\times32$ tori for the nearest-neighbor Hubbard model ($t'/t=0$), comparing HFPS \cite{Roth2025}, Transformer-BF \cite{Gu2026nc}, SCALE and ACE-BF \cite{Gu2026prx}, $16\times16$ PBC variational VAFQMC \cite{Sorella2023,Wu2024,Chen2026} and twist-averaged thermodynamic-limit (TDL) $\tau\to\infty$ Sorella VAFQMC \cite{Sorella2023}, variational ($D=16$) and $D\to\infty$ extrapolated fPEPS, NN-fPEPS ($D=4$), BCS backflow, and Pfaffian backflow \cite{Chen2026} with our multi-reference ($K=4$) variational curve and same-manifold $\tau^{-3}\to 0$ extrapolations ($K=4$ on $16\times16$, $K=1$ on $32\times32$; open vermilion stars).
(\textbf{B}) Energy per site $E/N$ for the cuprate Hubbard model ($t'/t=-0.2$) across $16\times16$, $24\times24$, and $32\times32$, comparing HFPS \cite{Roth2025}, SCALE and ACE-BF \cite{Gu2026prx}, symmetry-preserving BCS backflow \cite{Rende2026}, and NN-fPEPS ($D=4$) and Pfaffian backflow \cite{Chen2026} with our multi-reference variational benchmarks ($K=8$ on $16\times16$, $K=4$ on $32\times32$) and same-manifold $\tau^{-3}\to 0$ extrapolations ($K=1$ on $16\times16$, $K=4$ on $32\times32$; open vermilion stars).}
\label{fig:bench}
\end{figure}

Figure~\ref{fig:bench} and Table~\ref{tab:bench} benchmark the ground-state energies of our multi-reference MR-nVAFQMC wavefunctions at $\delta=1/8$ and $U/t=8$ across periodic $16\times16$ (256-site) and $32\times32$ (1,024-site) square tori. Throughout, we separate fixed-checkpoint variational upper bounds from empirical Sorella projection-depth extrapolations $E(\tau)/N=E_\infty/N+a\tau^{-3}$ \cite{Sorella2023} (marked ``not variational'' in Table~\ref{tab:bench} and shown as open stars in Fig.~\ref{fig:bench}), alongside both variational and extrapolated literature benchmarks. For the periodic $16\times16$ VAFQMC benchmark, $E/N=-0.75865(3)$ is the Sorella--Tocchio result reported in the V-score dataset of Wu \emph{et al.} \cite{Wu2024} and reproduced in Table S4 of Chen \emph{et al.} \cite{Chen2026}; Ref.~\cite{Sorella2023} describes the underlying method and its twist-averaged TDL extrapolation.

Across both $16\times16$ and $32\times32$ tori, MR-nVAFQMC reaches variational energies competitive with the most accurate neural- and tensor-network simulations reported to date (Fig.~\ref{fig:bench} and Table~\ref{tab:bench}). In the nearest-neighbor model ($t'/t=0$; Fig.~\ref{fig:bench}A), our $K=4$ states ($\tau=1.40$ on $16\times16$, $\tau=1.31$ on $32\times32$) match or lie slightly below BCS backflow, Sorella VAFQMC, Pfaffian backflow, ACE backflow, and $D=4$ NN-fPEPS within statistical uncertainty while lying well below unprojected SCALE and variational $D=16$ fPEPS \cite{Chen2026,Sorella2023,Wu2024,Gu2026prx}; extrapolating $E(\tau)/N$ in $\tau^{-3}$ on $16\times16$ ($-0.7604(32)$) further agrees closely with both $16\times16$ $D\to\infty$ extrapolated fPEPS ($-0.7602(14)$) \cite{Chen2026} and Sorella's twist-averaged 2D thermodynamic-limit $\tau\to\infty$ estimate ($-0.7605(1)$) \cite{Sorella2023}. In the frustrated cuprate model ($t'/t=-0.2$; Fig.~\ref{fig:bench}B), our $K=8$ ($\tau=1.45$) checkpoint on $16\times16$ and $K=4$ ($\tau=1.26$) checkpoint on $32\times32$ likewise match or lie slightly below Pfaffian backflow, symmetry-preserving BCS backflow, and $D=4$ NN-fPEPS within statistical error \cite{Chen2026,Rende2026}, demonstrating that particle-conserving multi-determinant neural auxiliary-field states access the same milli-$t$ low-energy regime without geminal pairing initialization.

\begin{table}[ht!]
\centering
\caption{\textbf{Best MR-nVAFQMC ground-state energies per site (this work) on large 2D square tori at $\delta=1/8$ ($U/t=8$) compared with leading published benchmarks.} For each periodic $L\times L$ torus and $t'/t$, our best multi-reference strictly variational state and same-manifold $\tau^{-3}\to 0$ extrapolation (marked ``not variational'') plotted in Fig.~\ref{fig:bench} are listed alongside the lowest published periodic-torus benchmarks \cite{Chen2026,Gu2026prx,Rende2026,Sorella2023,Wu2024} [for reference, Sorella's twist-averaged 2D thermodynamic-limit $\tau\to\infty$ extrapolation at $t'=0$ is $-0.7605(1)$ \cite{Sorella2023}]. Numbers in parentheses denote 1 s.e.\ statistical uncertainty in the last displayed digit(s) [e.g., $-0.7589(19)$ denotes $-0.7589\pm 0.0019$]; $|G|=8$ denotes projection onto lattice reflections ($R_x,R_y$) and global spin flip ($D_2\times\mathbb Z_2^{\mathrm{spin}}$), while $|G|=64,128$ additionally include commensurate stripe translations (Section~\ref{sec:method}).}
\label{tab:bench}
\footnotesize
\begin{tabular}{llr}
\toprule
Lattice, $t'/t$ & Method & $E/N$ ($t$) \\
\midrule
$16\times 16$, $0$ & \textbf{MR-nVAFQMC (this work)} ($K=4$, $|G|=8$, $\tau\to\infty$ extrap., not variational) & $-0.7604(32)$ \\
 & fPEPS, $D\to\infty$ extrap.\ (Chen 2026, not variational) & $-0.7602(14)$ \\
 & \textbf{MR-nVAFQMC (this work)} ($K=4$, $\tau=1.40$, $|G|=128$) & $-0.7589(19)$ \\
 & BCS backflow (Chen 2026) & $-0.75878(2)$ \\
 & VAFQMC (Sorella 2023; via Wu 2024, Chen 2026) & $-0.75865(3)$ \\
 & NN-fPEPS, $D=4$ (Chen 2026) & $-0.75851(2)$ \\
\addlinespace[3pt]
$16\times 16$, $-0.2$ & \textbf{nVAFQMC (this work)} ($K=1$, $|G|=64$, $\tau\to\infty$ extrap., not variational) & $-0.7482(19)$ \\
 & \textbf{MR-nVAFQMC (this work)} ($K=8$, $\tau=1.45$, $|G|=8$) & $-0.7466(38)$ \\
 & Pfaffian backflow (Chen 2026) & $-0.74533(1)$ \\
 & NN-fPEPS, $D=4$ (Chen 2026) & $-0.74467(2)$ \\
 & SBP (Rende 2026) & $-0.74411(1)$ \\
\addlinespace[3pt]
$32\times 32$, $0$ & \textbf{nVAFQMC (this work)} ($K=1$, $|G|=8$, $\tau\to\infty$ extrap., not variational) & $-0.7581(17)$ \\
 & \textbf{MR-nVAFQMC (this work)} ($K=4$, $\tau=1.31$, $|G|=8$) & $-0.7577(26)$ \\
 & NN-fPEPS, $D=4$ (Chen 2026) & $-0.75677(1)$ \\
 & BCS backflow (Chen 2026) & $-0.75617(1)$ \\
 & SCALE det. backflow (Gu 2026) \cite{Gu2026prx} & $-0.7501$ \\
\addlinespace[3pt]
$32\times 32$, $-0.2$ & \textbf{MR-nVAFQMC (this work)} ($K=4$, $|G|=8$, $\tau\to\infty$ extrap., not variational) & $-0.7494(35)$ \\
 & \textbf{MR-nVAFQMC (this work)} ($K=4$, $\tau=1.26$, $|G|=8$) & $-0.7476(37)$ \\
 & Pfaffian backflow (Chen 2026) & $-0.74569(1)$ \\
 & NN-fPEPS, $D=4$ (Chen 2026) & $-0.74257(2)$ \\
\bottomrule
\end{tabular}
\end{table}

\subsection{Directional pairing decay and exploratory transferred-parameter scaling}
\label{sec:scaling}

\begin{figure}[ht!]
\centering
\includegraphics[width=0.96\textwidth]{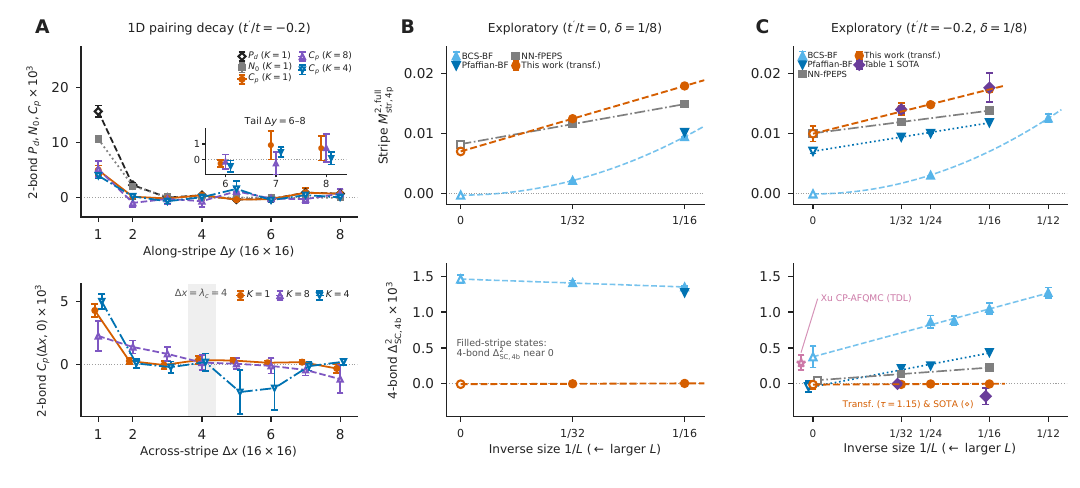}
\caption{\textbf{Directional decay of connected $d$-wave pairing on $16\times16$ and exploratory transferred-parameter scaling up to $32\times32$ at $\delta=1/8$.}
(\textbf{A}) Directional 1D two-bond cuts on $16\times16$ ($t'/t=-0.2$). Top: along-stripe bare $P_d(0,\Delta y)$ (open diamonds), Wick background $N_0(0,\Delta y)$ (grey squares) for $K=1$ ($\tau=1.52$), and connected vertex $\Cp(0,\Delta y)=P_d-N_0$ for $K=1,4,8$ (inset: $\Delta y=6$--$8$). Bottom: across-stripe $\Cp(\Delta x,0)$ ($\Delta x=\lambda_c=4$ shaded).
(\textbf{B},\textbf{C}) Exploratory fixed-depth transferred-parameter scaling ($1/L\to 0$) at (\textbf{B}) $t'/t=0$ ($K=1,\tau=1.16$; $L=16,32$) and (\textbf{C}) $t'/t=-0.2$ ($K=4,\tau=1.15$; $L=16,24,32$), comparing full-$\mathrm{SU}(2)$ stripe spin order $\Mfull$ (top) and the exact four-bond site-centered singlet distant-window pairing $\Dfour\times10^3$ [Eq.~(\ref{eq:pair_op}), matching Ref.~\cite{Chen2026}; bottom] of \textbf{This work (transf.)} (vermilion circles and dashed guides; $L=24,32$ parameters transferred from $16\times16$ without per-size re-optimization) together with unified same-checkpoint \textbf{Table 1 SOTA} evaluations on $16\times16$ ($K=8,\tau=1.45$) and $32\times32$ ($K=4,\tau=1.26$) (purple diamonds in \textbf{C}) against NN-fPEPS, Pfaffian backflow, and BCS backflow \cite{Chen2026,Rende2026} as well as pinned-field CP-AFQMC \cite{Xu2024} (open star). Error bars denote 1 s.e.\ [omitted for Ref.~\cite{Chen2026} points whose uncertainties lie below the quoted $3\times10^{-4}$ plot-unit threshold; the Ref.~\cite{Xu2024} star shows the asymmetric $1\sigma$ interval propagated from $\Delta_d=0.017(3)$].}
\label{fig:pair}
\end{figure}

We next examine how connected $d$-wave pairing correlations decay spatially on $16\times16$ and how spin and pairing observables behave in an exploratory fixed-depth series up to $32\times32$ (Fig.~\ref{fig:pair}). Along the stripe direction on $16\times16$ ($t'/t=-0.2$; Fig.~\ref{fig:pair}A, top), the bare two-bond correlator $P_d(0,\Delta y)$ at short distance is dominated by the disconnected one-body Wick contribution $N_0(0,\Delta y)$, underscoring the need to subtract $N_0$. The resulting connected intra-stripe vertex $\Cp(0,\Delta y)=P_d-N_0$ is clearly positive at nearest-neighbor separation ($\Delta y=1$) across $K=1,4,8$, but drops sharply already by $\Delta y=2$---where $K=1,4$ lie near zero within $1\sigma$ and $K=8$ fluctuates slightly negative within uncertainty---and remains statistically indistinguishable from zero over $\Delta y=6$--$8$ (Fig.~\ref{fig:pair}A, inset). Perpendicular to the stripes (Fig.~\ref{fig:pair}A, bottom), $\Cp(\Delta x,0)$ decays even more rapidly, becoming indistinguishable from zero within a single charge-stripe period ($\Delta x=\lambda_c=4$) and beyond. Deepening projection on $16\times16$ up to $\tau=1.45$--$1.52$ closes the energy offset relative to Pfaffian backflow while leaving both stripe spin order and distant-window pairing unchanged within $1\sigma$, and doping scans into $\delta=1/6$ and $2/9$ on $12\times12$ similarly enhance quasi-1D along-stripe $\Cp(0,1)$ while transverse and distant-window pairing remain unresolved.

To compare distant-window pairing directly with Chen \emph{et al.} \cite{Chen2026} across commensurate square tori up to $32\times32$, Fig.~\ref{fig:pair}B,C evaluates the full-$\mathrm{SU}(2)$ four-peak stripe spin order $\Mfull$ together with the exact four-bond site-centered singlet pairing order parameter $\Dfour(L)$ matching Eq.~(2) of Ref.~\cite{Chen2026} (Section~\ref{sec:method}). Two caveats govern the interpretation of the common-depth multi-size series and motivate labeling Fig.~\ref{fig:pair}B,C as \emph{exploratory}. First, the fixed-depth series ($\tau=1.16$ at $t'=0$; $\tau=1.15$ at $t'/t=-0.2$) uses shallower wavefunctions at higher variational energies than the deeper $\tau=1.26$--$1.45$ energy-benchmark checkpoints of Table~\ref{tab:bench} [for instance, on $32\times32$ at $t'/t=-0.2$, the $\tau=1.15$ evaluation gives $E/N=-0.736(6)$ compared with $-0.7476(37)$ for the $\tau=1.26$ checkpoint in Table~\ref{tab:bench}]. Second, the $24\times24$ and $32\times32$ wavefunctions in the common-depth series are constructed by tiling and transferring parameters optimized on $16\times16$ without independent variational re-optimization on the larger tori ($\mathrm{steps}=0$ on $32\times32$), and under $G=D_2\times\mathbb Z_2^{\mathrm{spin}}$ ($|G|=8$) these large-torus states remain frame-pinned ($A_4>0$). To remove checkpoint mismatch directly at $L=16$ and $L=32$, we additionally perform unified same-checkpoint joint evaluations of energy, $\Mfull$, and four-bond $\Dfour$ on the exact Table~\ref{tab:bench} SOTA wavefunctions ($16\times16$, $K=8,\tau=1.45$; $32\times32$, $K=4,\tau=1.26$; purple diamonds in Fig.~\ref{fig:pair}C).

Within both the common-depth transferred series and the unified Table~\ref{tab:bench} SOTA checkpoints, $\Mfull(L)$ retains pronounced period-16 ($t'=0$; Fig.~\ref{fig:pair}B, top) and period-8 ($t'/t=-0.2$; Fig.~\ref{fig:pair}C, top) stripe spin order up to $32\times32$, comparable in magnitude to $D=4$ NN-fPEPS and Pfaffian backflow \cite{Chen2026}, whereas $C_{4v}$-symmetric BCS backflow suppresses stripe magnetism toward a nearly uniform state \cite{Rende2026,Chen2026}. In the four-bond pairing channel (Fig.~\ref{fig:pair}B,C, bottom), $\Dfour(L)$ is compatible with zero at every torus size in the transferred series [giving a three-point linear $1/L$ intercept of $(-0.01\pm0.06)\times10^{-3}$ at $t'/t=-0.2$] and remains statistically indistinguishable from zero on the exact Table~\ref{tab:bench} SOTA checkpoints on $16\times16$ and $32\times32$ (purple diamonds); likewise, the maximum site-centered $d_{x^2-y^2}$ Bloch eigenvalue density $\max_{\mathbf Q}\tilde\Gamma_d(\mathbf Q)/N$ across all momenta $\mathbf Q$---including uniform $(0,0)$ and PDW $(\pi/4,0),(\pi/2,0)$---scales downward as $\mathcal O(1/N)$ from $8\times8$ to $32\times32$. By comparison, while $C_{4v}$-symmetric BCS backflow extrapolates to a sizable uniform order parameter \cite{Rende2026,Chen2026}, the lower-energy striped wavefunctions of Chen \emph{et al.} \cite{Chen2026} retain positive finite-size pairing on $16\times16$--$32\times32$ that extrapolates to very weak or fit-window-sensitive superconductivity in the thermodynamic limit [$\Delta_{\mathrm{SC}}=0.007(2)$ for $D=4$ NN-fPEPS, and $0.010(3)$ from $L=24,32$ versus a $1\sigma$ bound $\Delta_{\mathrm{SC}}<0.006$ from $L=16,24,32$ for Pfaffian backflow]. Because $\Delta_{\mathrm{SC}}\approx 0.007$ corresponds to $\Delta_{\mathrm{SC}}^2\approx 0.05\times10^{-3}$, comparable to our statistical uncertainty, our results neither resolve long-range pairing nor statistically exclude weak superconductivity at the NN-fPEPS scale; establishing fully converged ground-state scaling across all sizes will still require independent per-size variational re-optimization at matched deep $\tau$ across $L=16,24,32$.

\section{Method}
\label{sec:method}

\subsection{Multi-reference neural VAFQMC (MR-nVAFQMC) wavefunction}

Building on the general strategy of superposing nonorthogonal, symmetry-projected determinants \cite{JimenezHoyos2013}, we construct the multi-reference, symmetry-restored neural VAFQMC (MR-nVAFQMC) wavefunction
\begin{equation}
|\Psi(\theta,\mathbf c)\rangle=\sum_{k=1}^{K}c_k\,\hat P_G\hat P_\theta(\tau)|\Phi_k\rangle,
\label{eq:ansatz}
\end{equation}
where $\{|\Phi_k\rangle\}_{k=1}^{K}$ is a pool of inequivalent canonical particle-conserving Slater determinants spanning candidate stripe periods and shifted domain-wall configurations, $\hat P_G=|G|^{-1}\sum_{g\in G}\hat U_g$ projects onto a discrete lattice and spin-flip symmetry group $G$ of order $|G|$, and $\hat P_\theta(\tau)$ is a shared parameterized auxiliary-field correlation projector of imaginary-time depth $\tau$, extending single-reference VAFQMC \cite{Sorella2023} to coherent multi-determinant superpositions. Applying symmetric Suzuki--Trotter factorization and discrete Hubbard--Stratonovich decoupling across $L_\tau$ slices writes $\hat P_\theta(\tau)=\sum_{\mathbf x}p_\theta(\mathbf x)\hat B_\theta(\mathbf x)$, where $\mathbf x\in\{\pm1\}^{L_\tau N}$ denotes the auxiliary Ising fields and $\hat B_\theta(\mathbf x)$ is a product of spin-factored one-body exponentials. Beyond tunable Trotter slice weights, we incorporate a neural network in the auxiliary-field sector that takes the auxiliary-field configuration $\mathbf x$ as input and modulates the auxiliary-field path weights and effective one-body slice couplings shared across all $K$ reference determinants, increasing expressive flexibility while keeping every field configuration an exact one-body Slater-determinant propagation. On $64$-site ($16\times4$, $8\times8$) and $16\times16$ tori we also employ stripe or full space-group projections ($|G|=64,128,256,512$), whereas for the $16\times16$ ($t'/t=-0.2$, $K=8$) and $32\times32$ multi-reference benchmarks in Table~\ref{tab:bench} we use the eight-element group $G=D_2\times\mathbb Z_2^{\mathrm{spin}}$ ($|G|=8$) generated by lattice reflections $R_x,R_y$ and global spin flip $\uparrow\leftrightarrow\downarrow$, because summing many off-diagonal translations at large $\tau$ increases computational cost and sign fluctuations while the sub-extensive projection gain $\Delta E_{\mathrm{symm}}/N\sim\mathcal O(1/N)$ becomes small on $256$--$1,024$-site tori where shifted stripe sectors are represented through the multi-reference pool and translation-averaged observables. Whereas each individual unprojected Slater determinant $|\Phi_k\rangle$ carries no explicit BCS or Pfaffian pairing field and has zero connected two-body pairing ($\Cp=0$), symmetry restoration $\hat P_G$, multi-determinant interference ($K>1$), and the neural auxiliary-field projector $\hat P_\theta(\tau)$ together generate non-Gaussian spin, charge, and pairing correlations without geminal pairing initialization. Superposition weights $\mathbf c$ ($c_1\equiv 1$), reference orbitals, and projector parameters $\theta$ are optimized jointly by damped stochastic reconfiguration, followed by frozen-checkpoint sign-reweighted Markov-chain evaluation \cite{Sorella2023} with statistical uncertainties estimated by coarse-grained block jackknife over independent Markov chains after thermalization burn-in.

\subsection{Spin, charge, and connected pairing observables}

From the equal-time spin and density correlators $\langle\hat{\mathbf S}_i\cdot\hat{\mathbf S}_j\rangle=\langle\hat S_i^z\hat S_j^z\rangle+\tfrac{1}{2}\langle\hat S_i^+\hat S_j^-+\hat S_i^-\hat S_j^+\rangle$ and $\langle\hat n_i\hat n_j\rangle$ ($\hat n_i=\hat n_{i\uparrow}+\hat n_{i\downarrow}$), longitudinal spin structure factors and the full-$\mathrm{SU}(2)$ four-peak stripe magnetic order parameter (matching Ref.~\cite{Chen2026}) are obtained from
\begin{equation}
S_z(\mathbf q)=\frac{1}{N}\sum_{i,j}e^{i\mathbf q\cdot(\mathbf r_i-\mathbf r_j)}\langle\hat S_i^z\hat S_j^z\rangle,
\qquad
\Mfull=\frac{1}{4N}\sum_{\alpha=1}^{4} S(\mathbf q_\alpha),
\label{eq:spin_op}
\end{equation}
where $S(\mathbf q)=N^{-1}\sum_{i,j}e^{i\mathbf q\cdot(\mathbf r_i-\mathbf r_j)}\langle\hat{\mathbf S}_i\cdot\hat{\mathbf S}_j\rangle$ and $\mathbf q_\alpha\in\{(\pi\pm 2\pi\eps,\pi),(\pi,\pi\pm 2\pi\eps)\}$ with spin incommensurability $\eps=1/\lambda_s$. Static 1D hole-density modulation $\delta(x)=1-L_y^{-1}\sum_y\langle\hat n_{x,y}\rangle$ at commensurate stripe period $\lambda_c$ (wavevector $q_x=2\pi/\lambda_c$, e.g., $A_4$ at $q_x=\pi/2$) and translation-invariant connected charge fluctuations are quantified by
\begin{equation}
A_{\lambda_c}=\frac{2}{L_x}\left|\sum_{x=0}^{L_x-1}\delta(x)\,e^{-2\pi ix/\lambda_c}\right|,
\qquad
N_c(\mathbf q)=\frac{1}{N}\sum_{i,j}e^{i\mathbf q\cdot(\mathbf r_i-\mathbf r_j)}\bigl(\langle\hat n_i\hat n_j\rangle-\langle\hat n_i\rangle\langle\hat n_j\rangle\bigr),
\label{eq:charge_op}
\end{equation}
satisfying $N_c(\mathbf 0)=0$ identically in the canonical ensemble.

With normalized nearest-neighbor singlet creation operator $\hat b^\dagger_{ij}=(\hat c^\dagger_{i\uparrow}\hat c^\dagger_{j\downarrow}-\hat c^\dagger_{i\downarrow}\hat c^\dagger_{j\uparrow})/\sqrt{2}$ and $d_{x^2-y^2}$ bond signs $h_{\pm\hat x}=+1$, $h_{\pm\hat y}=-1$, we distinguish two standard $d$-wave pair operators: the two-bond link-origin operator $\hat\Delta^\dagger_{2\mathrm{b},\mathbf r}$ used for short-range vertices and 1D directional cuts (Figs.~\ref{fig:phase}B, \ref{fig:mech}B, and \ref{fig:pair}A), and the four-bond site-centered operator $\hat\Delta^\dagger_{4\mathrm{b},\mathbf r}$ matching Eq.~(2) of Ref.~\cite{Chen2026} used for distant-window comparisons in Fig.~\ref{fig:pair}B,C,
\begin{equation}
\hat\Delta^\dagger_{2\mathrm{b},\mathbf r}=\frac{1}{2}\sum_{\boldsymbol\eta\in\{+\hat x,+\hat y\}}h_{\boldsymbol\eta}\,\hat b^\dagger_{\mathbf r,\mathbf r+\boldsymbol\eta},
\qquad
\hat\Delta^\dagger_{4\mathrm{b},\mathbf r}=\frac{1}{4}\sum_{\boldsymbol\eta\in\{\pm\hat x,\pm\hat y\}}h_{\boldsymbol\eta}\,\hat b^\dagger_{\mathbf r,\mathbf r+\boldsymbol\eta}.
\label{eq:pair_op}
\end{equation}
From the bare pair correlator $P_d(\mathbf r)=N^{-1}\sum_{\mathbf r_0}\langle\hat\Delta^\dagger_{\mathbf r_0}\hat\Delta_{\mathbf r_0+\mathbf r}\rangle$ built from bond-singlet amplitudes $\langle\hat b_b^\dagger\hat b_{b'}\rangle$ [$b=(i,j)$, $b'=(k,l)$], subtracting the disconnected normal Wick contraction built from the full state's symmetrized equal-time one-body Green's function $\bar G_{\sigma;\,ij}=\langle\hat c^\dagger_{i\sigma}\hat c_{j\sigma}\rangle$,
\begin{equation}
\Cp(\mathbf r)=P_d(\mathbf r)-N_0(\mathbf r),
\qquad
N_{0,\,bb'}=\tfrac{1}{2}\bigl(\bar G_{\uparrow;\,ik}\bar G_{\downarrow;\,jl}+\bar G_{\uparrow;\,il}\bar G_{\downarrow;\,jk}+\bar G_{\downarrow;\,ik}\bar G_{\uparrow;\,jl}+\bar G_{\downarrow;\,il}\bar G_{\uparrow;\,jk}\bigr),
\label{eq:cp}
\end{equation}
isolates genuine connected two-body pairing in either convention \cite{Zhang1997,Chen2026} [used throughout Figs.~\ref{fig:mech} and \ref{fig:pair}; the broad doping scan in Fig.~\ref{fig:phase}B uses a uniform archive transition-$N_0$ subtraction across all points that agrees within statistical uncertainty with Eq.~(\ref{eq:cp}), e.g., $\Cp(0,1)=(+4.14\pm0.47)\times10^{-3}$ versus $(+4.35\pm0.61)\times10^{-3}$ on $16\times4$ at $\delta=1/8$]. From the two-bond vertex we form the cumulative short-range sum $\chivtx{4}=\sum_{0<|\mathbf r|\le 4}\Cp^{(2\mathrm{b})}(\mathbf r)$, while from the four-bond vertex we evaluate the 2D distant-window average matching Ref.~\cite{Chen2026},
\begin{equation}
\Dfour(L)=\frac{1}{M_L}\sum_{|\mathbf r|\ge L/(2\sqrt{2})}\Cp^{(4\mathrm{b})}(\mathbf r),
\label{eq:dsc}
\end{equation}
over the $M_L=159$, $355$, and $623$ periodic minimum-image displacements outside $L/(2\sqrt{2})$ on $L=16$, $24$, and $32$. On $64$-site tori, we also construct the full $2N\times2N$ bond-singlet matrix $\Gamma_{bb'}=\langle\hat b_b^\dagger\hat b_{b'}\rangle$ \cite{Yang1962} and suppress positive noisy-matrix diagonalization bias via the 50/50 split-block cross-validated Rayleigh quotient $\lambda_1^{\mathrm{CV}}=\tfrac{1}{2}\bigl(\mathbf v_{1,A}^\dagger\Gamma_B\mathbf v_{1,A}+\mathbf v_{1,B}^\dagger\Gamma_A\mathbf v_{1,B}\bigr)$, which provides a conservative (downward-biased in expectation relative to $\lambda_1(\Gamma)$, up to finite-sample ratio bias) estimate of the leading pair-matrix weight.

\section{Discussion and Conclusion}
\label{sec:conclusion}

Using MR-nVAFQMC wavefunctions built from particle-conserving Slater determinants without explicit pairing fields---at variational energies matching or improving upon leading neural- and tensor-network benchmarks up to $32\times32$ sites---our single-layer calculations establish a clear separation between intra-stripe local pair formation and inter-stripe phase coherence in the 2D $t$--$t'$--$U$ Hubbard model. Negative next-nearest-neighbor hopping $t'/t=-0.2$ switches the $\delta=1/8$ state from filled ($\lambda_c=8$) to half-filled ($\lambda_c=4,\lambda_s=8$) antiphase stripes obeying the Yamada relation $\eps=\delta$, stabilized by transverse kinetic delocalization against an opposing Hubbard repulsion penalty. Crucially, each half-filled domain wall already hosts robust short-range $d$-wave singlet pairing: the connected intra-wall vertex is enhanced more than fivefold over the filled stripes at $t'=0$, absent on the electron-doped N\'eel branch, peaking on the overdoped side near $\delta=3/16$ where period-eight spin stripe order softens, and unchanged within uncertainty at $\delta=1/8$ whether the walls are pinned or coherently delocalized across translated stripe sectors so that static charge modulation drops threefold.

Comparing these strictly single-layer results with the experimental $1/8$ anomaly in $\mathrm{La}_{1.875}\mathrm{Ba}_{0.125}\mathrm{CuO}_4$ highlights both an encouraging microscopic agreement and an instructive open tension \cite{Moodenbaugh1988,Li2007,Berg2007,Fradkin2015}. Experimentally, bulk 3D superconductivity at $x=1/8$ occurs at only $T_c\approx 4\text{ K}$ \cite{Li2007}, whereas each individual $\mathrm{CuO}_2$ layer exhibits 2D superconducting fluctuations setting in near the spin-stripe ordering scale ($\sim 40\text{ K}$) and reaches a 2D zero-resistance state near $16\text{ K}$; the suppression of 3D $T_c$ arises from interlayer Josephson decoupling, widely attributed to an intra-plane period-eight PDW that alternates sign between adjacent stripes and rotates by $90^\circ$ from layer to layer \cite{Li2007,Berg2007,Fradkin2015}. On the local scale, our finding that half-filled $\lambda_c=4$ domain walls intrinsically form short-range $d$-wave pairs---with uniform $\mathbf q=(0,0)$ and period-eight PDW $\mathbf q=(\pi/4,0)$ modes nearly degenerate in $64$-site pair-matrix spectra because inter-wall Josephson coupling across the antiphase spin domain is very weak---captures the microscopic origin of intra-wall pairing and near-degenerate $0$- versus $\pi$-phase inter-stripe channels. On the macroscopic 2D scale, however, our single-layer states show transverse pair correlations decaying within one stripe spacing ($\Delta x=4$) and resolve no long-range 2D superconducting order up to $32\times32$, in contrast to the 2D planar superconductivity observed below $16\text{ K}$ in $\mathrm{La}_{1.875}\mathrm{Ba}_{0.125}\mathrm{CuO}_4$ \cite{Li2007} and consistent with the weak or statistically unresolved 2D pairing in lower-energy striped Pfaffian and NN-fPEPS wavefunctions \cite{Chen2026}.

Three concrete possibilities can reconcile this tension and define a clear roadmap for future work. First, a weak 2D intra-plane condensate at the scale corresponding to $T_{\mathrm{BKT}}\approx 16\text{ K}\sim 4\times10^{-3}t$---such as the small amplitude $\Delta_{\mathrm{SC}}=0.007(2)$ ($\Delta_{\mathrm{SC}}^2\approx 5\times10^{-5}$) suggested by $D=4$ NN-fPEPS \cite{Chen2026}---lies at the statistical noise floor of our present distant-window estimator, and the spatially uniform average $\Dfour(L)$ cancels any sign-alternating $\mathbf q=(\pi/4,0)$ PDW component. Second, although unified same-checkpoint evaluations on the lowest-energy Table~\ref{tab:bench} wavefunctions at $16\times16$ and $32\times32$ confirm strong stripe spin order with unresolved four-bond $\Dfour$, the common-depth multi-size series in Fig.~\ref{fig:pair}B,C still relies on transferred $16\times16$ parameters on $24\times24$ and $32\times32$ under frame-pinned $D_2\times\mathbb Z_2^{\mathrm{spin}}$ symmetry ($A_4>0$). Independent per-size variational re-optimization at matched deep $\tau$, combined with direct superfluid-stiffness and translation-restored multi-sector scaling \cite{Kohn1964,Scalapino1993}, will test whether full large-torus relaxation resolves weak 2D uniform or PDW coherence. Third, if converged ground states of the pure single-band 2D $t$--$t'$--$U$ model at $\delta=1/8$ remain devoid of long-range 2D order, then stabilizing planar coherence at commensurate $1/8$ lock-in requires ingredients beyond the minimal single-band Hubbard Hamiltonian---such as multi-orbital $\mathrm{CuO}_2$ physics, longer-range interactions, or low-temperature-tetragonal lattice couplings \cite{Berg2007,Fradkin2015,Keimer2015}---or doping away from $1/8$ lock-in into the $\delta\gtrsim 3/16$ regime where spin stripes soften.

\begingroup
\raggedright

\endgroup

\end{document}